\documentclass[aps,amsmath,reprint,amssymb,superscriptaddress]{revtex4-2}

\usepackage{dcolumn}
\usepackage{graphicx}
\usepackage{color}
\usepackage{amsmath}

\begin{document}
%
\title{Visualization of optical polarization transfer to\\ photoelectron spin vector emitted from the spin-orbit coupled surface state}
\author{Kenta~Kuroda}
\email{kuroken224@issp.u-tokyo.ac.jp}
\affiliation{Institute for Solid State Physics (ISSP), University of Tokyo, Kashiwa, Chiba 277-8581, Japan}
\author{Koichiro~Yaji}
\email{YAJI.Koichiro@nims.go.jp}
\affiliation{Research Center for Advanced Measurement and Characterization, National Institute for Materials Science (NIMS), Tsukuba, Ibaraki 305-0003, Japan}
\thanks{These two authors contributed equally.}
\author{Ryo~Noguchi}
\author{Ayumi~Harasawa}
\affiliation{Institute for Solid State Physics (ISSP), University of Tokyo, Kashiwa, Chiba 277-8581, Japan}
\author{Shik~Shin}
\affiliation{Institute for Solid State Physics (ISSP), University of Tokyo, Kashiwa, Chiba 277-8581, Japan}
\affiliation{Office of University Professor, The University of Tokyo, Chiba 277-8581, Japan}
\author{Takeshi~Kondo}
\author{Fumio~Komori}
\affiliation{Institute for Solid State Physics (ISSP), University of Tokyo, Kashiwa, Chiba 277-8581, Japan}
\date{\today}
%
\begin{abstract}                                                   %
Similar to light polarization that is selected by a superposition of optical basis, electron spin direction can be controlled through a superposition of spin basis.  
We investigate such a spin interference occurring in photoemission of the spin-orbit coupled surface state in Bi$_2$Se$_3$ by using spin- and angle-resolved photoemission spectroscopy combined with laser light source (laser-SARPES).
Our laser-SARPES with three-dimensional spin detection and tunable laser polarization including elliptical and circular polarization enables us to directly visualize how the direction of the fully-polarized photoelectron spin changes according to the optical phase and orientation of the incident laser polarization.
By this advantage of our laser-SARPES, we demonstrate that such optical information can be projected to the three-dimensional spin vector of the photoelectrons.
Our results, therefore, present a novel spin-polarized electron source permitting us to optically control the pure spin state pointing to the arbitrary direction.
\end{abstract}
\maketitle
%
%
%
Realization of spin-polarized electrons and their manipulation are the key goal in the field of spintronics~\cite{Manchon_nm2015}.
A promising avenue to achieve it is utilizing materials with strong spin-orbit couplings (SOC).
The most vivid examples are Rashba systems~\cite{Bihlmayer_NJP2015} and topological insulators~\cite{Hasan10rmp, Ando13jpsj}, at surfaces of which spin-polarized electrons emerge due to the SOC in collaboration with the lack of the inversion symmetry.
In addition, as a consequence of the strong SOC, there can be a mix of orbitals that couple to different helical-spin components in the wavefunctions~\cite{Zhang13prl,Cao13NaturePhys,Wissing_prl2014}.
This spin-orbit entanglement can be directly probed by spin- and angle-resolved photoemission spectroscopy~\cite{Okuda_2017,Dil_2019} (SARPES), confirmed in various systems such as Bi$_2$Se$_3$~\cite{Zhu14prl, Xie2014naturecom,Kuroda_prb2016}, W(110)~\cite{Miyamoto_prb2016}, Bi(111)~\cite{Yaji_NatComm2017}, BiTeI~\cite{Maass_NC2016} and Bi/Ag(111)~\cite{Noguchi_prb2017}.
Since the light polarization can select the orbital part of the wavefunctions~\cite{Cao13NaturePhys}, this entanglement allows us to optically control the electron spin, leading to opto-spintronic functions~\cite{Mciver2012naturenano,Gmitra_prb2015}.

\begin{figure}[t!]
\begin{center}
\includegraphics[width=0.92\columnwidth]{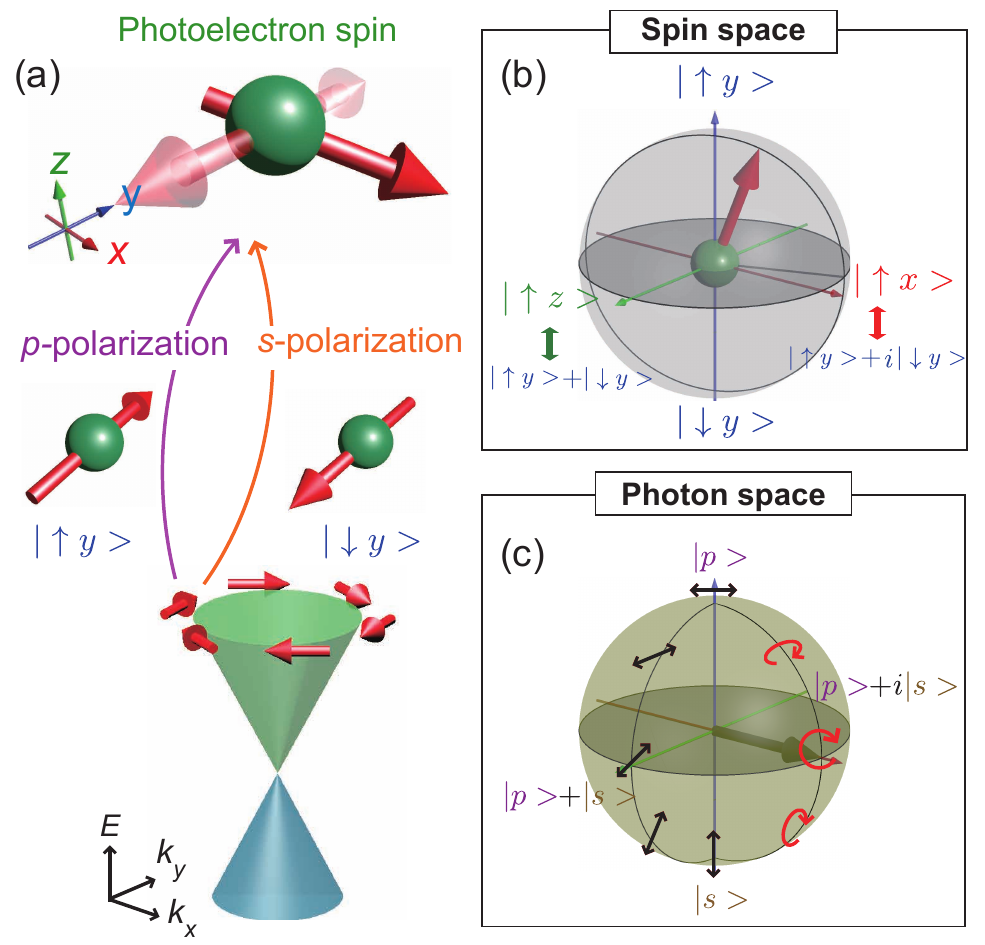}
\caption{
(a) Superposition of up-spin and down-spin in photoelectron state excited from the Dirac surface-state of Bi$_{2}$Se$_{3}$~\cite{Kuroda_prb2016}.
(b) Bloch sphere to display the resulting photoelectron spin.
Top (bottom) of the sphere indicates the up-spin (down-spin) along the initial helical spin components $y$-axis, which is selectively excited by the $p$-polarized ($s$-polarized) light.
The equator reflects the other spin components, determined by the coherent photoemission process~\cite{Yaji_NatComm2017,Kuroda_prb2016}.
(c) The circular and linear polarization plotted on the sphere (namely Poincar{\'e} sphere).
Let us use (top) $p$- and (bottom) $s$-polarization as a basis of the photon space.
}
\label{fig1}
\end{center}
\end{figure}
\begin{figure*}[t]
\begin{center}
\includegraphics[width=\textwidth]{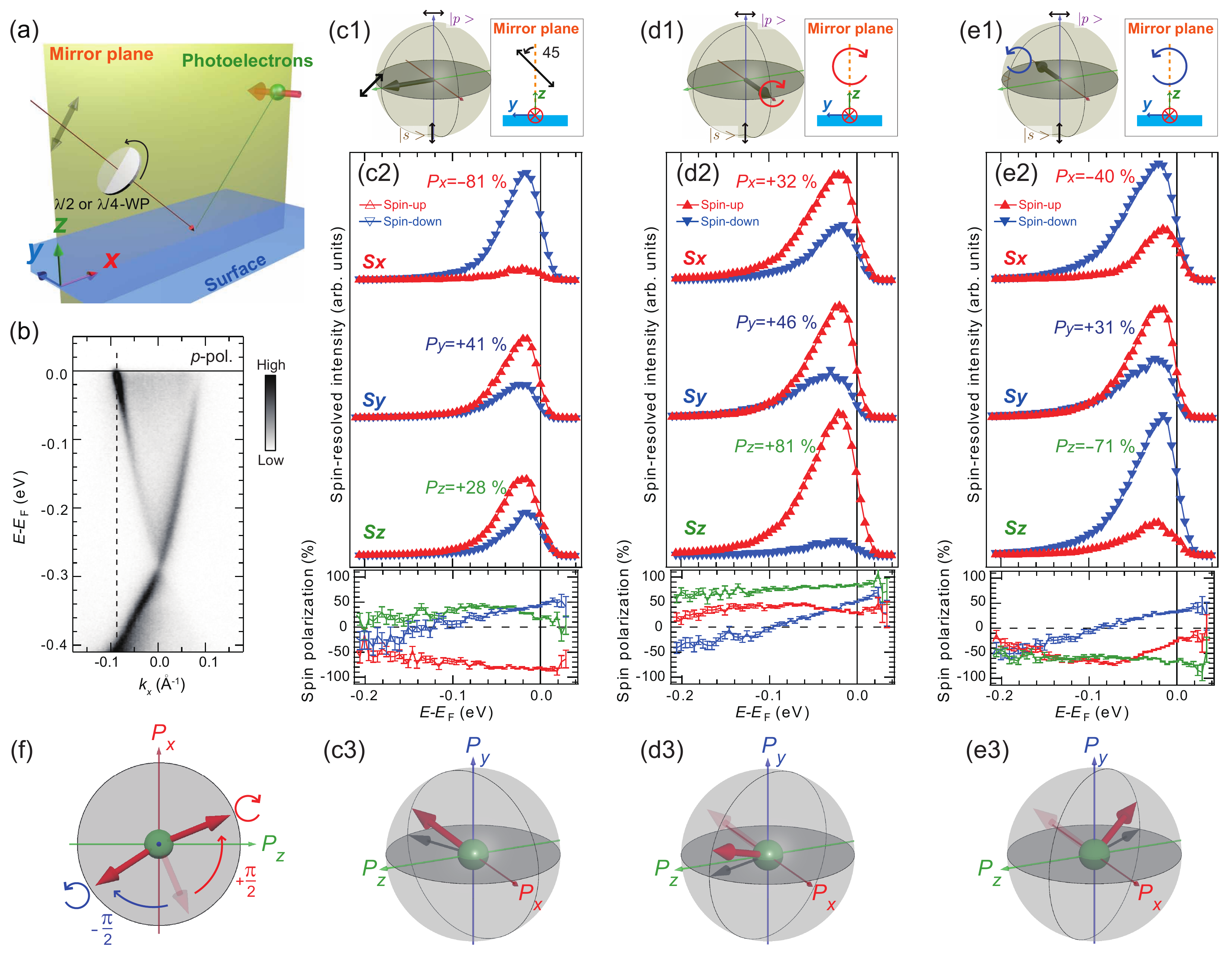}
\caption[]{(a) Experimental configuration where the incidence plane of the $p$-polarized light and the detection plane match the mirror plane of the crystal ($xz$-plane)~\cite{Kuroda_prb2016,Yaji_NatComm2017}.
The linear and circular polarization are controlled by the $\lambda$/2 and $\lambda$/4 waveplate (WP).
(b) ARPES map of the Dirac surface state in Bi$_2$Se$_3$(111) along the $\bar{\Gamma}$-$\bar{\rm{M}}$ line.
(c1-e1) The linear and circular polarization utilized for laser-ARPES plotted on the Poincar{\'e} sphere.
(c2-e2) Spin-resolved energy distribution curves at an emission angle denoted by the dashed line in (b), and the corresponding spin polarizations along $x$, $y$ and $z$ axes, $P_{x, y, z}$.
(c3-e3) The photoelectron spin vectors plotted on the Bloch sphere, which are experimentally determined by $P_{x, y, z}$ at $E-E_{\rm{F}}=-$0.02~eV.
The projected spin vectors onto the $P_z$-$P_x$ plane are denoted by black arrows.
(f) Comparison of these spin directions on the $P_z$-$P_x$ plane.
The spin obtained for the tilted linear polarization (c3) is displayed as the transparent arrow.
}
\label{fig2}
\end{center}
\end{figure*}
By previous SARPES studies on the surface state of Bi$_2$Se$_3$~\cite{Zhu14prl, Xie2014naturecom,Kuroda_prb2016}, it was demonstrated that $p$- or $s$-polarized light can selectively excite the fully spin-polarized photoelectrons with either spin-up or spin-down from spinor fields of the wavefunction.
Moreover, by using a $tilted$ linear polarization with both $p$- and $s$-components, the both spin states are excited simultaneously and interfere in photoelectron states [Fig.~\ref{fig1}(a)], which can be seen as a rotation of the spin orientation from the initial helical-spin state~\cite{Kuroda_prb2016}.
While the above experiments greatly facilitate an optical spin manipulation by a proper light polarization~\cite{Park_prl2012,Joswiak13NaturePhys}, a directional control of the electron spin pointing to the arbitrary direction has not been achieved yet.
To realize it, understanding the photoelectron spin excitation not only for the linear polarization but also for circular~\cite{Miyamoto2018,Joswiak13NaturePhys,Barriga14prx} and elliptical polarization is important, and such a comprehensive measurement also requires a three-dimensional (3D) spin detection to fully trace the rotation of the photoelectron spin but has never been performed so far.

In this Letter, we use SARPES combined with laser light source (laser-SARPES) and the 3D spin detection, and design the experiment with variable laser polarization on the spin-orbit entangled surface state of Bi$_2$Se$_3$.
Since the spin rotation is reflected by the interference of a two-level system $|{\uparrow}>+e^{i\phi}|{\downarrow}>$ [Fig.~\ref{fig1}(a)], we here use a Bloch sphere to display it [Fig.~\ref{fig1}(b)]: the poles correspond to the spin polarization direction of the initial state ($e.x.$ along $y$, $P_{y}$) and the equator represents the other spin components ($P_{x,z}$) emerging from the interference.
Our laser-SARPES with 3D spin detection enables us to obtain full information about the photoelectron spin vector and directly visualize how it changes together with the laser polarization, the optical orientation as well as the optical phase that all can be represented by Poincar{\'e} sphere [Fig.~\ref{fig1}(c)].
Our results unambiguously reveal that such optical information of the incident laser is fully projected to the 3D spin vector of the excited photoelectrons through the coherent photoemission process.

Laser-SARPES experiments were performed at ISSP, The University of Tokyo~\cite{Yaji_RSI2016,Kuroda_jove2018} utilizing a high-flux 6.994-eV laser source~\cite{ShimojimaJPSJ2015} and very low-energy electron diffraction (VLEED)-based spin polarimeters~\cite{Okuda_2017}.
Our spectrometer combined with the double VLEED detectors resolves the 3D spin polarization of the photoelectrons~\cite{OKUDA_201523}. 
In our experimental configuration, the mirror plane of the surface coincides with the plane of the incidence [Fig.~\ref{fig2}(a)], which is a necessary condition for symmetry arguments of the spin interference~\cite{Kuroda_prb2016,Yaji_NatComm2017}. 
The linear and circular polarization for the incident laser were controlled by the $\lambda$/2 and $\lambda$/4 plates made by MgF$_2$ crystal (Kougakugiken Corp.).
The degree of the polarization of the light is 97~$\%$~\cite{Yaji_prb2018}. 
During the measurement, the sample temperature was kept at 20~K, and instrumental energy and momentum resolutions were set below 20~meV and 0.7$^{\circ}$, respectively.
Single-crystal Bi$_{2}$Se$_{3}$ was cleaved $in\; situ$ by the Scotch tape at room temperature followed by a rapid cooling. 

Let us start with showing our laser-SARPES result for a tilted linear polarization [Fig.~\ref{fig2}(c1)]: a superposition of $p$- and $s$-polarization ($\epsilon_{p}$ and $\epsilon_{s}$, respectively).
Figure~\ref{fig2}(c2) represents the spin-resolved energy distribution curves (EDCs) with the corresponding spin polarizations $P_{x,y,z}$ obtained from the surface Dirac-cone at $-k_{\rm{F}}$ [dashed line in Fig.~\ref{fig2}(b)].
At the measured high-symmetry $k$-line along $\bar{\Gamma}$-$\bar{\rm{M}}$, only $P_{y}$ is allowed in the the initial state wavefunction according to the mirror symmetry~\cite{Henk_prb2003,Kuroda_prb2016}.
However, the tilted light polarization breaks the overall symmetry including the experimental geometry and thus allows $P_{x,z}$ in the photoelectron spin according to the interference of the initial state spins~\cite{Kobayashi_prb2017}.

Since our experimental geometry is appropriate for the spin selective excitation of $\epsilon_{p}$ and $\epsilon_{s}$, the measured $P_{x,y,z}$ through the spin interference can be expressed by a simple form~\cite{Kuroda_prb2016,Yaji_NatComm2017}:  
\begin{equation}
\begin{split}
P_{x}&={2sin\phi}\frac{|M_p||M_s|}{|M_p|^2+|M_s|^2}\\
P_{y}&=\frac{|M_p|^2-|M_s|^2}{|M_p|^2+|M_s|^2}\\
P_{z}&={2cos\phi}\frac{|M_p||M_s|}{|M_p|^2+|M_s|^2}
\label{eq1}
\end{split}
\end{equation}
where $M_p$ and $M_{s}$ are the dipole matrix elements for $\epsilon_{p}$ and $\epsilon_{s}$, respectively, and $\phi$ is a relative phase of these two matrix elements.
The photoelectron spin is thus depicted with pure two-spinors, $|M_p||{\uparrow}_{y}>+e^{i\phi}|M_s||{\downarrow}_{y}>$.
Accordingly, the spin interference and its key factor $\phi$ can be reflected by the $P_z$-$P_x$ plane (equatorial plane) of the Bloch sphere:
\begin{equation}
{\phi}=tan^{-1}\frac{P_{x}}{P_{z}}
\label{eq2}
\end{equation}
As seen in the spin orientation determined by the observed $P_{x,y,z}$ [Fig.~\ref{fig2}(c3)], $\phi$ is intrinsically present and previously determined to be $-\pi$/3 at the measured $k$ point~\cite{Kuroda_prb2016}.

As a heart of our interests, we now move to photoelectron spin excitations with the circular polarization that has a relative phase of $\pm\pi$/2 between $\epsilon_{p}$ and $\epsilon_{s}$ in the photon space [Figs.~\ref{fig2}(d1) and (e1)].
This helical polarization, therefore, should add $\pm\pi$/2 to a phase difference between $M_p$ and $M_s$, which can be reflected by a rotation of the photoelectron spin in the $P_z$-$P_x$ plane according to Eq.~(\ref{eq2}).
To examine the role of this helicity in the spin-rotation, full information about the photoelectron spin $P_{x,y,z}$ is necessary, which however has been lacking from the previous works on the surface state of Bi$_2$Se$_3$~\cite{Joswiak13NaturePhys, Barriga14prx}.

\begin{figure}[t]
\begin{center}
\includegraphics[width=\columnwidth]{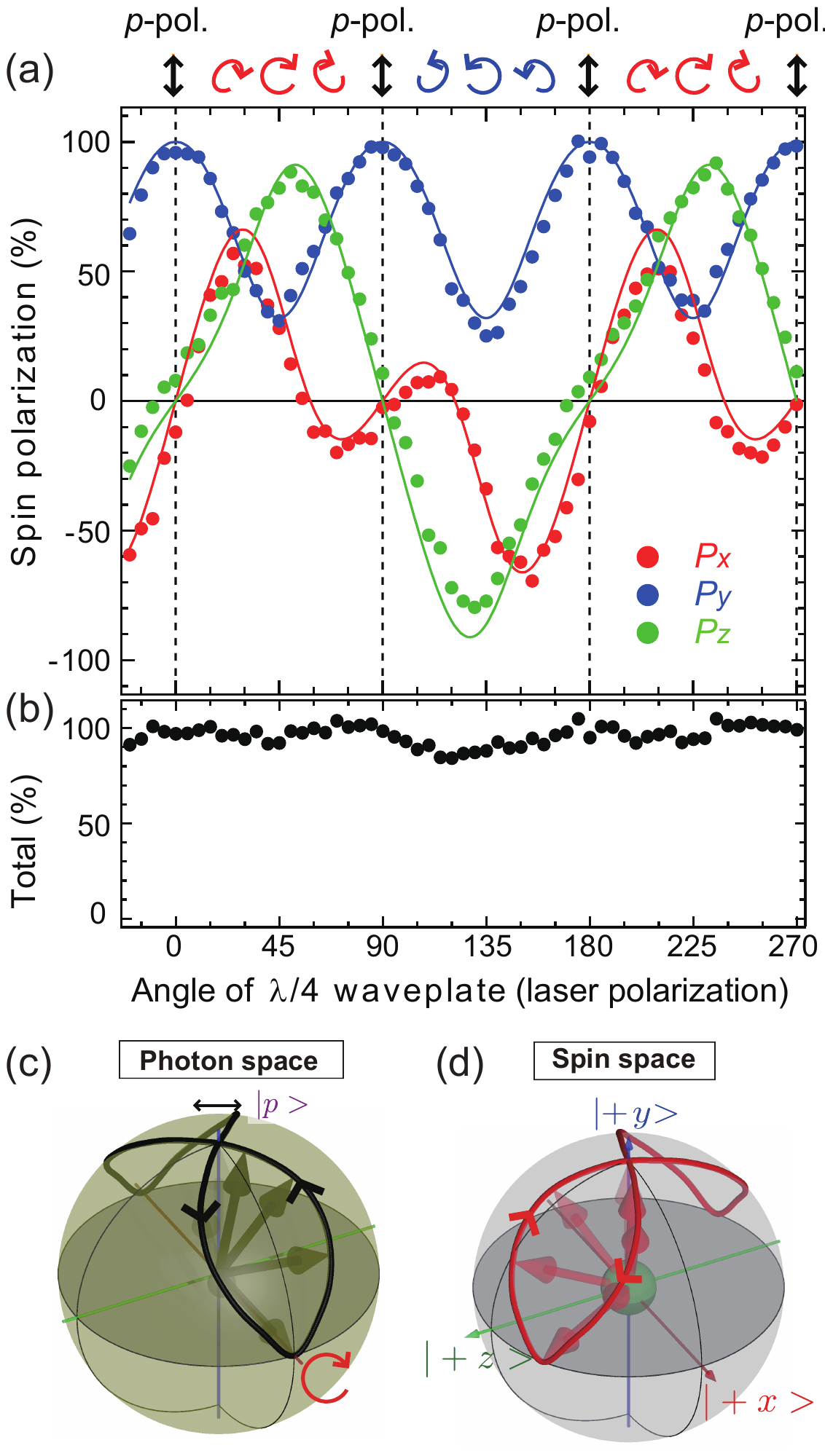}
\caption[]{(a) Plots of the observed $P_{x,y,z}$ as a function of the laser polarization.
Insets sketch the laser polarization simulated by Jones calculus for a rotation of $\lambda$/4-waveplate.
(b) The total spin polarization $P$=$\sqrt{P_{x}^{2}+P_{y}^{2}+P_{z}^{2}}$.
The solid curves are obtained by our simulation based on Eq.~(\ref{eq1}).
The details of the simulation are presented in our supplementary note.
(c) and (d) The photon space and the spin polarization direction of the photoelectrons.
Each trajectory is denoted by the bold lines.
}
\label{fig3}
\end{center}
\end{figure}

The impact of the helicity, revealed by our laser-SARPES with 3D spin detection, is displayed in the spin-resolved EDCs [Figs.~\ref{fig3}(d2) and (e2)]; $P_{x,z}$ significantly deviates from those for the linear polarization while the change of $P_{y}$ is negligibly small [Figs.~\ref{fig2}(c3-e3)].
Note that $P_{y}$ is insensitive to the optical phase because it relies only on the amplitude of $M_{p}$ and $M_{s}$ and does not include the complex information [see Eq.(\ref{eq1})].
In Fig.~\ref{fig2}(f), we present the determined spin orientation projected onto the $P_z$-$P_x$ plane.
Upon the circular polarization excitation, it rotates nearly $\pm\pi{/2}$ in the spin space with respect to that for the tilted linear polarization (the transparent arrow), corresponding to the helical state of the incident photon.
This Bloch sphere mapping by using 3D spin detection, thus, directly verifies that the complex information of the photon space [Figs.~\ref{fig2}(c1-e1)] can be transfered to the spin space [Figs.~\ref{fig2}(c3-e3)].

As the best advantage of utilizing the laser for SARPES, one can generate various laser polarization from linear to circular and even to be elliptical field just by changing a angle of the $\lambda$/4-waveplate.
In contrast to all previous SARPES experiments reported so far, our laser-SARPES affords to not only investigate circular polarization excitations but also fully trace this light-polarization evolution of all spin components $P_{x,y,z}$.
The data obtained by such a comprehensive experiment is presented in Fig.~\ref{fig3}.

Only $P_{y}$ is allowed for $\epsilon_{p}$ excitation and achieves +100$\%$ due to the selection rule of the initial spin-orbit wavefucntion~\cite{Kuroda_prb2016}.
By changing to the circular polarization, $P_{y}$ continuously decays to be +30$\%$, and instead both $|P_{x,z}|$ abruptly rises up.
Our data represents the overall oscillations of  $P_{x,y,z}$ as a function of the laser polarization.
One may find that $P_{x,z}$ ($P_{y}$) shows antisymmetric (symmetric) behavior for the different helicity ($\epsilon_{p}\pm{i}\epsilon_{s}$).
This can be elucidated with the mirror symmetry in our experimental geometry [Fig.~\ref{fig2}(a)].
Since the two helical polarizations can be transformed to each other by the mirror symmetry [insets of Figs.~\ref{fig2}(d1) and \ref{fig2}(e2)], one can obtain the following antisymmetric (symmetric) relation of $P_{x,z}$ ($P_{y}$): $P_{x,z}$(${\epsilon_{p}}+i{\epsilon_{s}}$)$\leftrightarrow-P_{x,z}$(${\epsilon_{p}}-i{\epsilon_{s}}$) and $P_{y}$(${\epsilon_{p}}+i{\epsilon_{s}}$)$\leftrightarrow{P_{y}}$(${\epsilon_{p}}-i{\epsilon_{s}}$).

The observed total spin $\sqrt{{P_{x}}^{2}+P_{y}^{2}+P_{z}^{2}}$ is nearly 100~$\%$ ($>$85~$\%$) for any light polarization [Fig.~\ref{fig3}(b)], which indicates that the spin system obtained by our laser-SARPES is considered as a coherent superposition of the two pure spin-level.  
Thanks to this simplicity, the laser-polarization evolution of $P_{x,y,z,}$ can be fully reproduced by the simple simulation considering for the quantum mechanics with Eq.~(\ref{eq1}).
There are only two parameters considered: the shape of the laser polarization and the intrinsic phase difference $\phi$ of $-\pi$/3 for the linear polarization~\cite{Kuroda_prb2016} (see supplementary text for the details).
Even with such a simple model, our simulation well clarifies the evolution of $P_{x,y,z,}$ [solid lines in Fig.~\ref{fig3}(a)].

By virtue of this good consistency, our laser-SARPES enables us to visualize the connection of the laser polarization and the photoelectron spin vector in Figs.~\ref{fig3}(c) and \ref{fig3}(d).
It is immediately found that their trajectories in the spheres guided by the bold lines show the similar feature.
Although this demonstration is performed only by rotating the $\lambda$/4 wavelate, it is enough to trace the evolutions of both the optical shape and its phase and validate the one-to-one correspondence between the laser polarization and the photoelectron spin vector.
Our result, therefore, manifests that one can control the spin pointing to any direction in three-dimension by a selection of the optical state.

The spin interference visualized by our laser-SARPES occurs during the coherent photoemission process.
However, the obtained $P_{x,y,z}$ should sensitively depend on experimental conditions, $e.x.$ photon energy, angle of light incidence~\cite{Dil_2019,Kobayashi_prb2017}, since incoherent superpositions of different channels may influence the photoemission, namely final state effect~\cite{Heinzmann_2012}.
Although its description, in general, requires one-step theory of photoemission~\cite{Braun_1996}, our spin system detected by laser-SARPES reflects rather simple situation with the pure two spin-state, which allows us to fully depict the laser polarization evolution of the resulting the 3D photoelectron spin. 
Moreover, circular polarization excitation has recently emerged as a powerful tool to generate and control photocurrents of highly spin-polarized electrons~\cite{Mciver2012naturenano,Photo_nm2015,Kuroda_prb2017}, activate properties of topological bands~\cite{Xu2018,Wang_Science2013} as well as valley polarization~\cite{Ye2017}.
The knowledge about the optical spin control of the electrons established by our laser-SARPES, therefore, should be critically important for future optspintronic applications.
%

In conclusion, we have investigated the photoelectron spin excitation from the surface state of Bi$_2$Se$_3$ by means of laser-SARPES combined with various laser polarization.
This measurement directly images how the direction of the photoelectron spin changes according to the optical orientation and its phase, and directly demonstrate that such an optical information can be fully transfered to the spin orientation of the photoelectrons through the spin interference during the photoemission process.
This work finds a pathway for optical control of the electron spin, which should play a role in opt-spintronic applications.

%
We gratefully acknowledge funding JSPS Grant-in-Aid for Scientific Research (B) through Projects No. 26287061, 19H02683, for Scientic Research (C) through Projects No. 18K03484, and for Young Scientists (B) through Projects No. 15K17675, 17K14319.
%

%

\end{document}